\documentclass[runningheads]{llncs}
\usepackage{graphicx}
\newcommand\bull{\noindent\hbox to\parindent{\footnotesize$\bullet$}}

\begin{document}
\title{Improving the Effective Utilization of Supercomputer Resources by Adding Low-Priority Containerized Jobs\thanks{The work was supported by the Russian Foundation for Basic Research grant 18-37-00502 ``Development and research of methods for increasing the performance of supercomputers based on job migration using container virtualization".}}
%
%
\author{Julia Dubenskaya\orcidID{0000-0002-2437-4600} \and Stanislav Polyakov\inst{*}\orcidID{0000-0002-8429-8478}}
\authorrunning{J. Dubenskaya, S. Polyakov}
%
\institute{Skobeltsyn Institute of Nuclear Physics, M.V.Lomonosov Moscow State University (SINP MSU), 1(2), Leninskie gory, GSP-1, Moscow 119991, Russia
\email{s.p.polyakov@gmail.com}}
\maketitle              
\begin{abstract}
We propose an approach to utilize idle computational resources of supercomputers. The idea is to maintain an additional queue of low-priority non-parallel jobs and execute them in containers, using container migration tools to break the execution down into separate intervals. We propose a container management system that can maintain this queue and interact with the supercomputer scheduler. We conducted a series of experiments simulating supercomputer scheduler and the proposed system. The experiments demonstrate that the proposed system increases the effective utilization of supercomputer resources under most of the conditions, in some cases significantly improving the performance.

\end{abstract}
\section{\label{section1} Introduction}
Data processing is one of the most important parts of the data life cycle. In contemporary practice, supercomputers are increasingly used to process large amounts of data, as well as to simulate various phenomena. In general, supercomputers cannot utilize all of the available computational resources at all times. In this paper, we focus on improving the performance of supercomputers by utilizing idle resources.

Modern supercomputers typically have hundreds or thousands of users who submit a broad range of different computing jobs. Managing these jobs is done by a scheduler which is a piece of software that maintains a job queue, assigns computational resources to jobs, and optimizes the performance of the supercomputer based on priorities such as maximizing the average load, minimizing the average time or the maximum time spent in the queue, or some combination of these or other metrics.

We focus on average load and closely related metrics. For many supercomputers, the average load is approximately 90\% and can be as low as 70\% \cite{Antonov,Zhumatiy}. It may be caused by insufficient number of submitted jobs (underload), unfitting job sizes (i.e. the jobs cannot fit together into a schedule without leaving some resources idle), inaccurate estimates of execution time used for planning the schedule in advance, different optimization priorities or restrictions, etc. 

One possible approach to increasing the average load of a supercomputer is adding low-priority jobs that can utilize idle computational resources. A possible source of such jobs are scientific experiments that generate petabytes of data that require primary processing. It is possible to split the data into fragments and process them independently. Another class of jobs required by scientific experiments is generating samples by Monte Carlo simulation. These jobs are available in effectively infinite numbers and can be executed sequentially and independently, thus they can utilize any available idle resources. An example of this approach is the use of idle resources of the Titan-2 supercomputer by the ATLAS project \cite{ATLAS}.

The implied assumption of this approach is that additional jobs should not have a significant negative effect on the performance parameters related to regular jobs. However, even if additional jobs are scheduled with a low priority, this can divert computational resources from executing regular jobs. The reason for this is that regular jobs require sufficient amount of resources to be available simultaneously, whereas additional jobs can start on any resources as soon as they become available.

A standard basic algorithm for many schedulers is a backfilling algorithm \cite{Lifka} that uses reservations to ensure that larger jobs are eventually scheduled and fills the schedule before the reservation time with jobs that will not interfere with the reservation.
Our experiments show that using this algorithm with reservations for regular jobs still does not prevent the diversion of computational resources from regular jobs in some cases, particularly for the additional jobs with large execution time (see section \ref{section4}).

This problem can be solved or partially solved with the help of some modifications of the scheduling algorithm. Another possible solution is to reduce the execution time of additional jobs, preferably in a way that is best suited for scheduling regular jobs, e.g. by making all concurrent additional jobs finish execution exactly before the reservation time. This solution can be implemented with the tools used in container virtualization.

Container virtualization is a method to isolate groups of processes from the rest of the system in their own user-space instances. Containers are similar to virtual machines but always use the same OS kernel as the host machine. Containers tend to be much more lightweight and take less time to start than virtual machines. 

One of the promising features supported by many container virtualization platforms is the checkpoint/restore in userspace (CRIU) mechanism. This mechanism allows one to freeze a running application and create a checkpoint, saving it as a collection of files. Then the files can be used to restore the application and run it exactly as it was during the time of the freeze, possibly in another instance of the operating system. In container virtualization systems, the CRIU mechanism enables the migration of containers. 

We propose an approach to increasing the load of supercomputers by adding low-priority queues of non-parallel jobs. Executing the jobs in containers allows the checkpoint mechanisms to be applied to them when necessary, saving the progress of execution and returning them to the queue. With this approach, the negative effect of additional queues on the performance parameters related to regular jobs can often be negligible.
The approach addresses the underload problem and to some degree the problems of unfitting job sizes and inaccurate execution time estimates.

We have presented a similar approach in \cite{Polyakov}. Further research has shown that we did not take into account several important considerations. Most importantly, we underestimated the time needed for checkpoint procedures. In this paper, we update our approach based on these considerations.
Similar idea was also proposed in \cite{Baranov} where a quasi scheduler was used to assign idle resources to additional jobs. Unlike regular jobs which require a specific amount of computational resources and execution time, and unlike additional jobs used in our approach which can have an arbitrary execution time, the jobs used by quasi scheduler request a specific execution time but are not restricted to a specific amount of computational resources.

\section{\label{section2} Container Checkpoints}

To test our idea, we chose the container platform Docker \cite{Docker}, since it is the most actively developing project in this area. We also use the fact that Docker containers are lightweight (compared to virtual machines), so a single server or computational node can simultaneously run multiple containers without noticeable decrease in performance.

For Docker, we performed a series of tests for its container migration functions and measured the time to create a checkpoint and the time to restore from a checkpoint. The parameters of the test machine are as follows: \newline
\bull total memory 16GB, swap 16GB; \newline
\bull processor: 16 cores Intel (R) Xeon (R) CPU E5620 @ 2.40GHz, architecture x86\_64; \newline
\bull operating system: CentOS Linux release 7.5.1804; \newline
\bull kernel version: 3.10.0-862.14.4.el7.x86\_64.

The measurements show that the time to create a checkpoint, as well as the time to restore from a checkpoint, have a normal distribution with a very small value of standard deviation. The average values of these parameters are highly correlated to the amount of random access memory (RAM) used by the container and almost do not depend on the CPU load.
We tested the time to create a checkpoint and the time to restore from a checkpoint for the containers using 1 MB, 100 MB, 200 MB, 400 MB, 800 MB, and 1.6 GB RAM. The average time to create a checkpoint is, respectively, 1.05, 5.45, 9.81, 19.6, 41.0, and 78.4 seconds. The average time to restore from a checkpoint is, respectively, 1.26, 5.0, 9.22, 17.1, 31.0, and 61.8 seconds. From the obtained results it can be concluded that the dependence on the amount of RAM used is close to linear in both cases. 

The experiments show that the jobs can be successfully saved and restored from the checkpoints and these procedures take reasonable time for the purposes of our proposed approach.

\section{\label{section3} The Architecture of the Proposed Container Management System}

We assume that our system will interact with a supercomputer scheduler that organizes the execution of queued jobs on a collection of computational resources. We will call the minimal amount of resources the scheduler can assign a computational node. We also assume that all computational nodes controlled by the same scheduler are equivalent.

We propose a container management system for supercomputers that manages the execution of containerized jobs on the computational nodes allotted by the supercomputer scheduler. The system has two components: a master program running in the shared memory and a local manager running on the computational nodes.

The master program submits jobs consisting of an instance of the local manager to the low-priority queue of the scheduler. It also maintains its own queue of non-parallel jobs to be started in containers, including both new jobs submitted by users and unfinished jobs returned by instances of the local manager.

Instances of the local manager are started on the allotted computational nodes by the scheduler. Each instance creates containers, pulls jobs from the master program queue, and starts them or restores them from checkpoints in the containers. Before the allotted time is over, the instance creates checkpoints for all the jobs it has started that are not finished yet and returns them to the master program queue.

Implementing the container management system does not require changes to the supercomputer scheduler if the scheduler supports an additional low-priority queue, because the scheduler does not need to make other distinctions between jobs submitted to it by the system and regular jobs. In this approach, the master program must request the execution time for each job it submits to the scheduler queue. 
It is possible to adjust the requested time separately for each job.
But without any way to know in advance how much time a job will spend in the low-priority queue, we do not see how this adjustment can be used to improve performance.
However, having the requested time set for all the container jobs can significantly reduce CPU time used by main queue jobs because the container jobs may gradually take over the nodes. We conducted a series of experiments (similar to those described in the next section) confirming that adding the infinite queue of one-hour non-parallel jobs generally decreases the average load by main queue jobs, sometimes by as much as several percent.

Our proposed solution to this problem is synchronized node release: all concurrent instances of the local manager exit at the same time which is published by the master program, instead of using the entire time allotted by the scheduler. We will call the time between synchronized node releases the synchronization frame.

If changes to the scheduler are allowed, several approaches can be implemented. The scheduler has more flexibility than the master program: it can synchronize the node release partially if there is no need for all of the nodes used by additional jobs, or refrain from starting additional jobs if the idle nodes can soon be assigned to a main queue job. The time allotted to an additional job can be modified when it is already running. Alternatively, the modified scheduler can let container jobs run without explicit time limits and stop them whenever the nodes are needed, if the local manager creates checkpoints periodically.

However, the appropriate changes to the scheduler depend on the specifics of the scheduling algorithm it uses. In the rest of the paper we consider the approach without modifying the scheduler as more general and easier to implement.

\section{\label{section4} Simulation Experiments}

We conducted two series of experiments in order to determine the potential increase in the efficient utilization of computational resources with the proposed approach. In the experiments we applied the EASY Backfill scheduling algorithm (\cite{Lifka}) to the job queues that were generated based on historical distribution of the job parameters (execution time and number of used nodes) from Lomonosov supercomputers.

In the first series, we assumed that there was sufficient number of jobs submitted to the main queue, so the computational nodes could only remain idle if all the jobs in the queue required more nodes or more time than was available. The experiments were conducted for the number of nodes ranging from 1024 to 4000 and determined the average load by main queue jobs, as well as the effective utilization of the nodes when the container management system was simulated. 

In the experiments with the number of nodes and job parameter distributions similar to those of Lomonosov-1 and Lomonosov-2 supercomputers the resulting average load was approximately 99.2\% and 97.1\%, respectively. This is significantly higher than the historical figures from Lomonosov-1 and Lomonosov-2 from 2016-2017 (92.3\% and 88.7\%, respectively, see \cite{Zhumatiy}). We assumed that the reason for this discrepancy was the underload, possibly in conjunction with administrative and user restrictions having similar effect of reducing the number of jobs available for scheduling. 

In our second series of experiments jobs were added to the main queue using the Poisson process with the submission rates chosen to approximate the historical underload. An effectively infinite additional job queue can obviously help solve the underload problem by itself. Therefore to measure the improvements resulting from our proposed approach the experiments with the additional queue were conducted both with and without simulating the container management system.

\subsection{Job Queues and Simulation Parameters}

Since scheduling simulation does not need to actually run any jobs, each job was represented by three parameters: the number of nodes it uses, the execution time, and the requested time. 
We used two types of queues, one with the number of nodes and execution time distribution similar to that of historical jobs on Lomonosov-1 supercomputer in 2018, and another one approximating Lomonosov-2 historical job parameter distribution in 2016--2017. We will call them L1 queue and L2 queue, respectively. L1 queue jobs had an average of 12.97 nodes with a standard deviation of 24.13 nodes, an average execution time of 400.6 minutes with a standard deviation of 979.8 minutes, an average size of 9479 node-minutes with a standard deviation of 40065 node-minutes. L2 queue jobs had an average of 4.209 nodes with a standard deviation of 6.765 nodes, an average execution time of 266.3 minutes with a standard deviation of 1332 minutes, an average size of 1450 node-minutes with a standard deviation of 16216 node-minutes. 
The requested time was generated for each job based on its actual execution time. Based on the observations of user estimates of execution time from \cite{Tsafrir} as well as the data and observations from Lomonosov supercomputers we chose a model with four cases: \newline
\bull accurate predictions: the requested time is equal to the execution time (also including the jobs that don't complete in the allotted time and have to be terminated by the scheduler); \newline
\bull moderate everestimations: the requested time is the least of the several round values (10 minutes, 30 minutes, 1 hour, 2 hours, 5 hours, 12 hours, 1 day, 3 days, 7 days, 15 days) greater than the execution time; \newline
\bull requesting the default time (1 day), unless the execution time is greater, and same as the previous case otherwise; \newline
\bull requesting the maximum allowed time (assumed 3 days for the queues based on Lomonosov-1 data and 15 days for the queues based on Lomonosov-2 data). \newline
Each case was chosen at random with a probability 1/4.

For the first series of the experiments, the main queue length was kept at 100 jobs with the new jobs added immediately to replace the ones that were scheduled. The second series of experiments were conducted using the Poisson process for adding jobs to the main queue. The parameters for the Poisson process were chosen in such a way that the average load of each of the simulated supercomputers with L1 and L2 queues (without additional jobs) was within 0.5\% of the historical average load of the respective supercomputer.

In the first series of experiments the schedules were made for 1024, 1500, 2000, 3000, and 4000 nodes (4000 and 1500 are the approximate numbers of nodes of the largest fragments of Lomonosov-1 and Lomonosov-2 supercomputers, respectively; 1024 is the maximum number of nodes that a single job is allowed to use). In the second series of experiments there were only two types of schedules: for 4000 nodes with L1 main queue, and for 1500 nodes with L2 main queue.

In each series of experiments the scheduling was made for L1 and L2 main queues with and without the additional queues of containerized jobs. We used the synchronized release approach for the containers with the synchronization frames of 30, 45, 60, 90, 120, and 180 minutes, and the additional 240 and 360 minutes in the second series. In the experiments from the second series where the non-containerized jobs were added, their execution time was 6, 12, 24, and 48 hours.

The duration of the scheduler slot was 1 minute. Each individual experiment simulated scheduling for 180 days. For each set of parameters (the main queue type, the number of nodes, the execution time of additional jobs or the synchronizaion frame if either applied) we conducted 50 experiments if the proposed container management system was simulated and 100 experiments otherwise.

\subsection{Results}

The first series of experiments demonstrates that both L1 and L2 queues have enough small-sized jobs to maintain very high average load without additional jobs if a sufficient number of jobs is kept in the queue at all times. The average load increases with the number of the available nodes and surpasses 99\% at 4000 nodes with both L1 and L2 queues. Interestingly, the average number of idle nodes changes only moderately depending on the total number of nodes: it ranges between 31.4 and 33.6 nodes for L1 queue and between 36.3 and 46.2 for L2 queue. 

Adding containerized jobs with release synchronization further increases the average load, at the cost of a small fraction of CPU time taken out of executing the main queue jobs. However, the added load is not equivalent to the load by regular jobs: the additional jobs have lower priority and use CPU time less efficiently because of the significant time needed to create checkpoints and restart from checkpoints. To make the comparison more meaningful we measure the effective utilization of computational resources instead of the load: $u = l - l_{\rm{aux}}$ where $l$ is the average load and $l_{\rm{aux}}$ is the average load by auxiliary checkpoint procedures. There are many variables that can influence the time required for auxiliary checkpoint procedures. For simplicity, we assume that every time a node executes containerized jobs, checkpoint procedures take 10 minutes out of the allotted time. To take into account the difference in priorities we separately calculate the average load by main queue jobs $l_m$. If the average load by main queue jobs is lower than the average load without the additional queue $l_{\rm{default}}$, we also calculate the trade-off factor
$F=\frac{u-l_m}{l_{\rm{default}}-l_m}$
(the ratio between the CPU time effectively used by additional jobs and the CPU time taken out of executing main queue jobs). If one can assign a numeric value to the relative importance of main queue jobs compared to additional jobs, then the proposed container management system should only be used if the expected trade-off factor is significantly higher than this value.

\begin{figure}
\includegraphics[width=\textwidth]{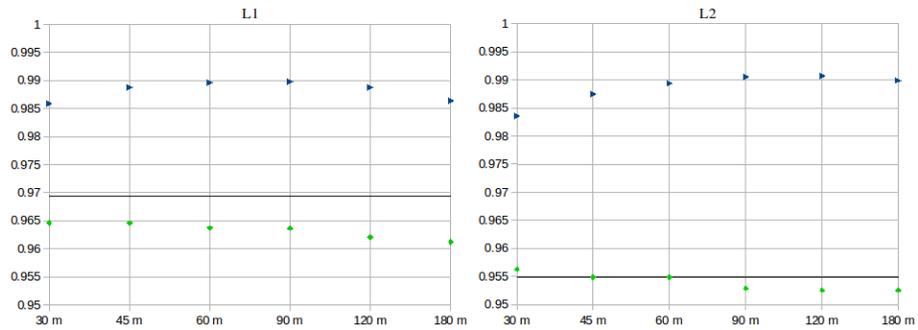}
\caption{Average load without additional jobs (black line), average load by main queue jobs (green rhombi), and effective utilization (blue triangles) for 1024 nodes, L1 (left) and L2 (right) main queues, synchronization frames from 30 to 180 minutes.}
\end{figure}
\begin{figure}
\includegraphics[width=\textwidth]{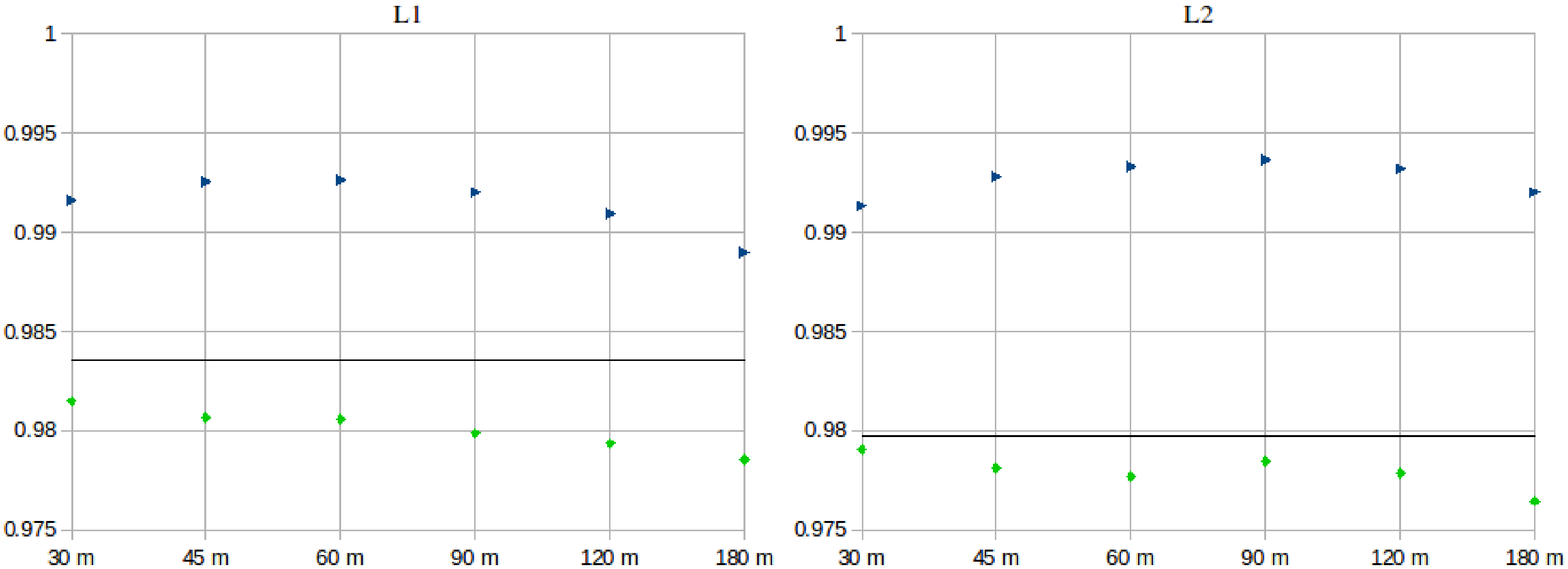}
\caption{Average load without additional jobs, average load by main queue jobs, and effective utilization for 2000 nodes.}
\end{figure}
\begin{figure}
\includegraphics[width=\textwidth]{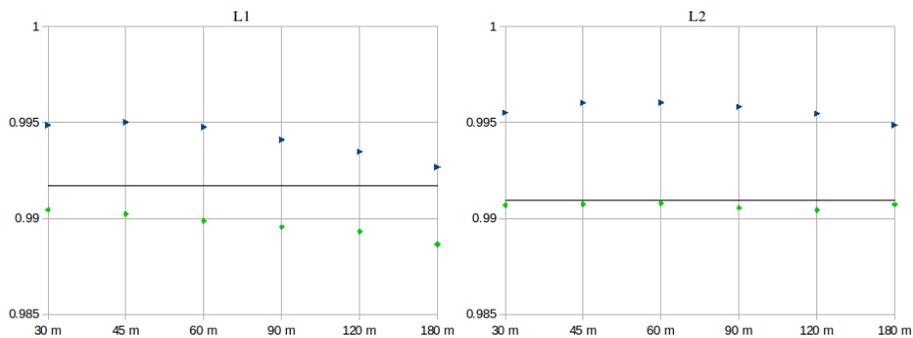}
\caption{Average load without additional jobs, average load by main queue jobs, and effective utilization for 4000 nodes.}
\end{figure}

The results for 1024, 2000, and 4000 nodes are presented on figures 1--3.
The maximum effective utilization over different synchronizaion frames is approximately 99.0\% and 99.1\% for L1 and L2 main queues with 1024 nodes, 99.5\% and 99.6\% with 4000 nodes, respectively.
The potential decrease in the average number of idle nodes or nodes performing auxiliary container procedures is 21 and 36.8 nodes for L1 and L2 main queues with 1024 nodes, 13.2 and 20.5 nodes for L1 and L2 main queues with 4000 nodes, respectively. Relative decrease is 3 and 4.9 times for L1 and L2 main queues with 1024 nodes, 1.7 and 2.3 times for L1 and L2 main queues with 4000 nodes, respectively. 

For L1 main queue, the trade-off factor is moderate and generally decreases with the increasing number of nodes and the synchronization frame. It never exceeds 5.1 and is below 3.5 for 4000 nodes. For L2 main queue, the trade-off factor is quite high: above 7.8 in all of the experiments with synchronization frames up to 120 minutes. For any given number of nodes there is a synchronization frame resulting in trade-off factor above 18.9.

\begin{figure}
\includegraphics[width=\textwidth]{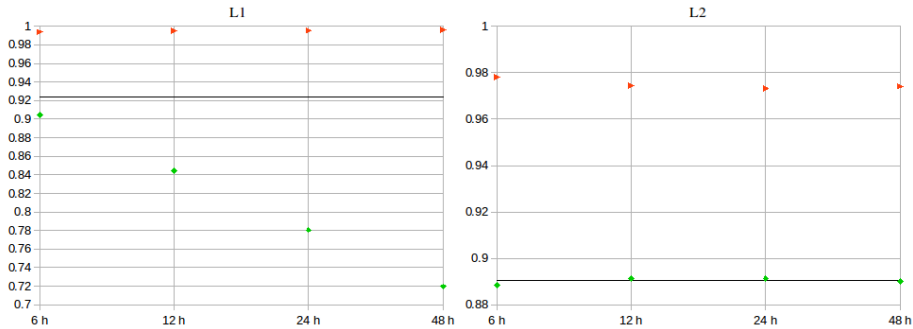}
\caption{Average load without additional jobs (black line), average load by main queue jobs (green rhombi), and average load (red triangles) for L1 main queue with 4000 nodes, $l_{\rm{default}}=0.924$ (left) and L2 main queue with 1500 nodes and $l_{\rm{default}}=0.8906$ (right), executing additional jobs without containers.}
\end{figure}

\begin{figure}
\includegraphics[width=\textwidth]{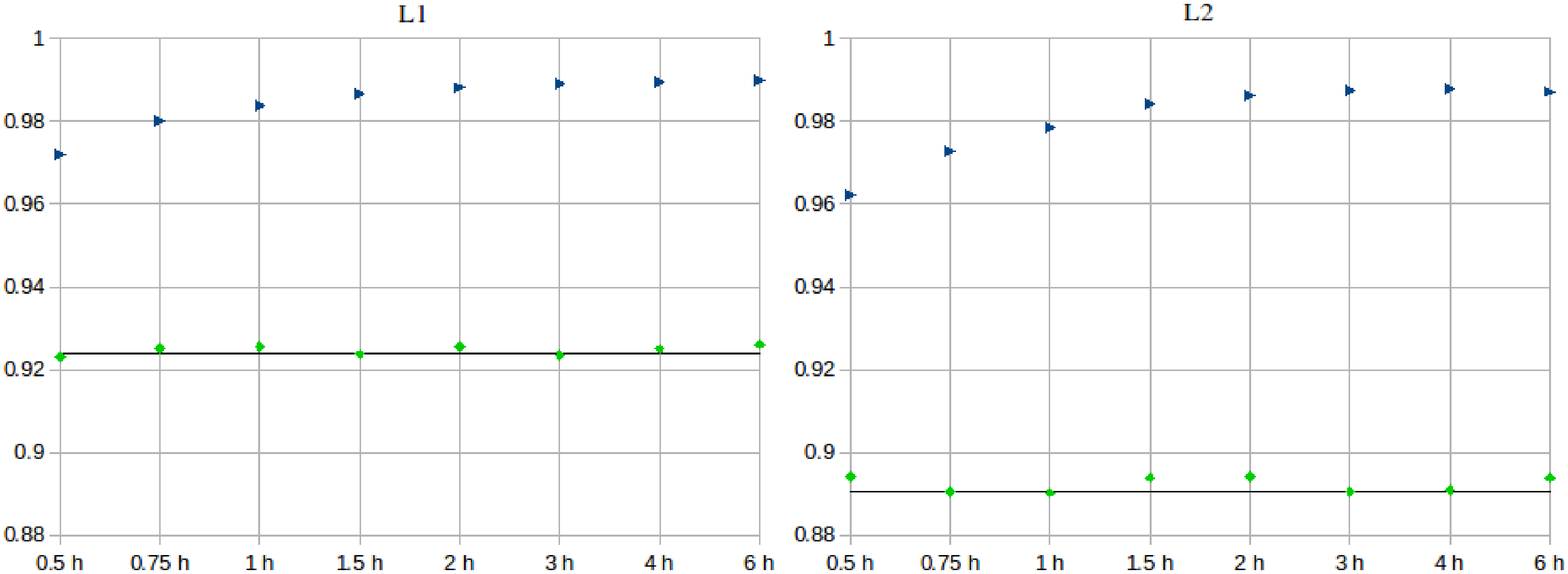}
\caption{Average load without additional jobs (black line), average load by main queue jobs (green rhombi), and effective utilization (blue triangles) for L1 main queue with 4000 nodes, $l_{\rm{default}}=0.924$ (left) and L2 main queue with 1500 nodes and $l_{\rm{default}}=0.8906$ (right), executing additional jobs in containers.}
\end{figure}

In the second series of the experiments adding low-priority queue of non-parallel jobs without containerization increases the average load to 99.4\%--99.6\% for L1 main queue (with 4000 nodes and $l_{\rm{default}}=0.924$) and to 97.3\%--97.8\% for L2 main queue (with 1500 nodes and $l_{\rm{default}}=0.8906$). The corresponding increase in the average number of busy nodes is 279 to 289 out of 4000 for L1 main queue and 124 to 131 out of 1500 for L2 main queue. However, there is a difference in the average load by main queue jobs: with L2 main queue it does not change significantly, and with L1 main queue it decreases significantly, resulting in the trade-off factor of 4.6 for 6-hour jobs and below 2 for longer jobs despite higher average load.

In the experiments with simulated container management system, synchronized node release can increase the average load by main queue jobs. The effective utilization generally increases with the synchronization frame. With L2 main queue it exceeds the average load without containers but with L1 main queue it is always lower than the respective average load. The maximum decrease in the number of idle nodes or nodes performing auxiliary container procedures is 263 out of 4000 nodes for L1 main queue and 146 out of 1500 nodes for L2 main queue. The results are presented on figures 4--5. 

In summary: \newline
\bull with L1 main queue without underload the effective utilization increase and the trade-off factor are not very high, although not necessarily low enough to rule out the use of our proposed system; \newline
\bull with L2 main queue without underload the effective utilization increase is higher and the trade-off factor can be very high, making this case a relatively good one to use the system; \newline
\bull with L1 main queue with underload the effective utilization increase is very high and the alternative approach of using the additional jobs without containerization can be unacceptable, making this case the best one to use the system; \newline
\bull with L2 main queue with underload the effective utilization increase is very high but most of it is due to the additional job queue itself. The benefits of containers are moderate but can justify the use of the system thanks to the lack of trade-off.

\section{Conclusion}

We proposed a system that manages the execution of non-parallel jobs in containers and maintains an additional low-priority queue for the supercomputer scheduler.
Our simulation experiments demonstrate that under some assumptions the system increases the effective utilization of computational resources. The increase is not always significant and sometimes the system diverts computational resources from executing regular jobs, but in many cases we expect the benefits to be considerable.

\subsubsection{Acknowledgements.} The work was supported by the Russian Foundation for Basic Research grant 18-37-00502 ``Development and research of methods for increasing the performance of supercomputers based on job migration using container virtualization". We would also like to thank Sergey Zhumatiy for the provided information and useful discussions.

%
%
%
%

\end{document}